\documentclass[
twocolumn,
  prl,
  showpacs,
  amsmath,
  amssymb,
  superscriptaddress,
  floatfix
]{revtex4}

\usepackage{graphicx}
\usepackage{color}
\usepackage{bm}
\usepackage{float}

\begin{document}
\title{Manifestation of electron-electron interaction in the
  magnetoresistance of graphene}

\author{Johannes Jobst}
%
\author{Daniel Waldmann}
\affiliation{Lehrstuhl f\"ur Angewandte Physik, Universit\"at
  Erlangen-N\"urnberg, 91058 Erlangen, Germany}
%

\author{Igor V.\ Gornyi}

\affiliation{
 Institut f\"ur Nanotechnologie, Karlsruhe Institute of Technology,
 76021 Karlsruhe, Germany
}
\affiliation{
 A.F.~Ioffe Physico-Technical Institute,
 194021 St.~Petersburg, Russia.
}

\author{Alexander D.\ Mirlin}

\affiliation{
 Institut f\"ur Nanotechnologie, Karlsruhe Institute of Technology,
 76021 Karlsruhe, Germany
}

\affiliation{
 Inst. f\"ur Theorie der kondensierten Materie,
 Karlsruhe Institute of Technology, 76128 Karlsruhe, Germany
}
\affiliation{
 Petersburg Nuclear Physics Institute,
 188300 St.~Petersburg, Russia.
}

\author{Heiko B. Weber}
\email{heiko.weber@physik.uni-erlangen.de}
\homepage{http://www.lap.physik.uni-erlangen.de/}
\affiliation{Lehrstuhl f\"ur Angewandte Physik, Universit\"at
  Erlangen-N\"urnberg, 91058 Erlangen, Germany}

\date{\today}


\date{\today}

\begin{abstract}
We investigate the magnetotransport in large area graphene Hall bars epitaxially grown
on silicon carbide. In the intermediate field regime between weak localization and Landau
quantization the observed temperature-dependent parabolic
magnetoresistivity (MR) is a manifestation of the electron-electron
interaction (EEI).
We can consistently describe the data with a model for diffusive (magneto)transport
that also includes magnetic-field dependent effects originating from ballistic time scales.
We find an excellent agreement between the experimentally
observed temperature dependence of MR and the theory of EEI in the diffusive regime.
We can further assign a temperature-driven crossover to the reduction of the
multiplet modes contributing to EEI from 7 to 3 due to intervalley scattering.
In addition, we find a temperature independent ballistic contribution to the MR
in classically strong magnetic fields.
\end{abstract}


\maketitle

Graphene is an outstanding electronic material with unique physical properties and with excellent prospects for technological applications
\cite{geim-rmp, novoselov-nobel, avouris-GHz, bae-roll}. The linear band
structure and the chiral nature of charge carriers result in a rich variety
of phenomena \cite{novoselov-QHE, kim-QHE, Klein-tunneling,
  savchenko-WL,neto-review} which can be largely described within a model of
non-interacting electrons.
Evidence for electron-electron interaction (EEI) in graphene, however, is
sparse. The most spectacular example is the occurrence of the
fractional quantum Hall effect and the insulating phase in strong
magnetic fields \cite{andrei-fractional, kim-fractional}.
Another line of research involves the investigation of EEI-induced
renormalization of the Dirac spectrum in graphene
\cite{Elias11,Siegel11}.

Furthermore EEI plays a prominent role in electron transport
also in the absence of quantizing magnetic field, when
the system is a Fermi liquid.
In particular, EEI leads to a quantum contribution to
zero-field resistivity, that in two dimensional electron gases (2DEG)
at sufficiently low temperatures $T$ displays a
logarithmic $T$ dependence \cite{altshuler}.
Logarithmic corrections may further stem from weak localization (WL), when observed in very low fields. At somewhat
higher fields  logarithmic $\rho(T)$ data were interpreted as EEI
\cite{camassel-EEI,shytov-EEI, lara-avila-EEI} or Kondo effect \cite{fuhrer-kondo}.
This indicates that an evaluation of the temperature dependence $\rho(T)$ alone is insufficient
for identification of the underlying mechanisms.
A way to unambiguously single out the EEI effect is to study the temperature-dependent
magnetoresistance (MR) of a 2DEG in non-quantizing magnetic fields, between weak fields where WL is
operative and strong fields where Landau quantization is fully
developed. The analysis of EEI effects in the magnetotransport,
which have been widely disregarded, is the scope of this paper.

For this investigation large sample dimensions are mandatory, as in small
flakes mesoscopic conductance fluctuations would add to the MR and the finite sample size sets a
low-energy cut-off to the EEI. This is the case for most experiments using exfoliated graphene.
Therefore we use epitaxial graphene on insulating silicon carbide (SiC) which is available with
excellent homogeneity on the wafer scale \cite{emtsev-natmat}.
Our special choice is quasi free-standing monolayer graphene (QFMLG)
\cite{riedl-QFMLG, speck-QFMLG} which represents
a hole-conducting graphene material with a charge carrier density of $n \approx 3\cdot10^{12}$\,cm$^{-2}$ and
a mobility $\mu$ of approximately 2000\,cm$^2$/Vs. In contrast to other epitaxial graphene
materials $n$ and $\mu$ are more or less independent of temperature due to a reduced coupling to
the substrate \cite{speck-QFMLG}. The chiral nature of charge carriers in QFMLG has been proven in previous
experiments \cite{waldmann-bottom}.

\begin{figure}[t]
    \includegraphics[width=1.00\columnwidth]{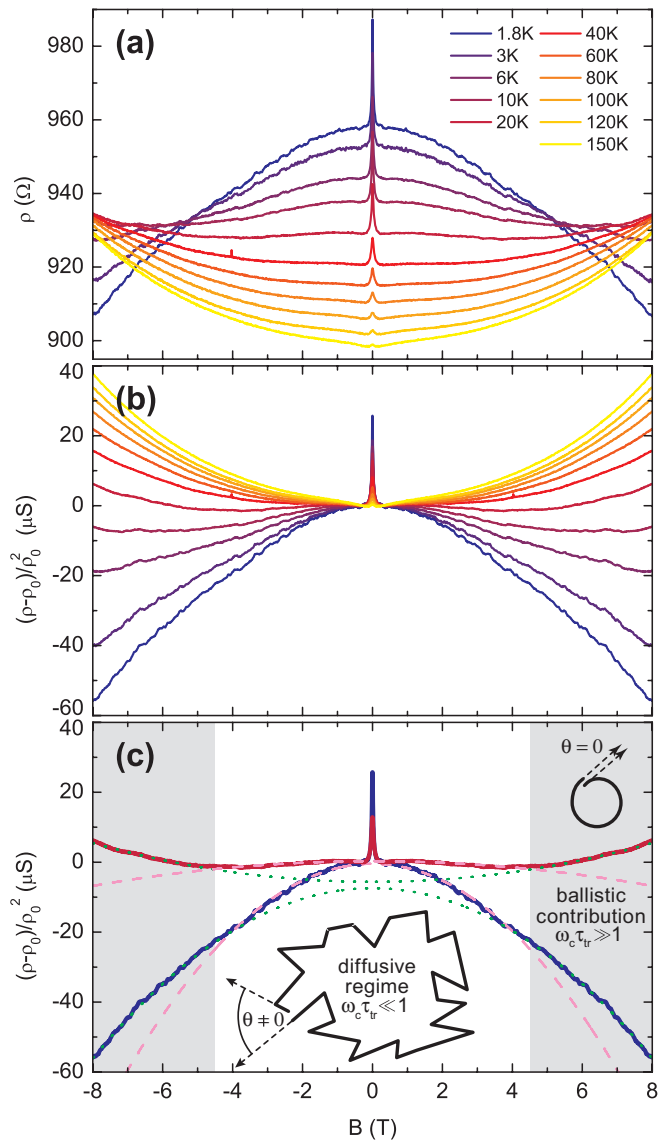}
    \caption{Evolution of magnetoresistivity with
          temperature. (a) The curvature changes sign with increasing
          temperature. Further, for temperatures $T \leq 10$\,K the
          curves cross at $B \approx 5$\,T; for
          higher temperatures the crossing point shifts towards a
          larger value  $B \approx 9$\,T. (b) Relative
          magnetoresistivity $(\rho-\rho_0)/\rho_0^2$.
          (c) Two parabolas $(\omega_{\text{c}}\tau_{\text{tr}})^2\frac{e^2}{2 \pi^2 \hbar}\cdot M$ are
          fitted to the data in (b) (shown representative for 1.8\,K and
          20\,K): one in the diffusive regime for weak fields
          $1\,\text{T}<|B|<4.5\,\text{T}$ (dashed) and one in the
          diffusive regime with ballistic contributions for
          $|B|>5\,\text{T}$ (dotted). Two sketches illustrate the
          different scattering mechanisms in both
          regimes. \label{fig:magnetoresistivity}}
\end{figure}

For the present study we prepared Hall bars by standard electron beam lithography \cite{jobst-QHE}
and measured the magnetoresistivity $\rho(B)$ in four-point geometry for various
temperatures $T$ using low-frequency lock-in technique. For all
$\rho(B,T)$ curves the magnetic field $B$ was swept
from -8 to 8\,T twice to confirm reproducibility. All measurements
were performed under cryogenic vacuum in order to reduce the effect of
adsorbates. In total we investigated four
samples of different size with very similar findings as concerned the EEI effects \cite{supplemental}. In this
paper we focus on one particular sample S1 (330\,$\mu$m long and 50\,$\mu$m wide channel)
which shows all observed phenomena in one consistent data set. Figure
\ref{fig:magnetoresistivity}(a) shows the magnetoresistivity raw data.
Obviously the Landau quantization regime is not
yet reached
(however, it has been observed in higher fields \cite{waldmann-bottom}).
This indicates that the damping of Shubnikov-de Haas oscillations in the sample is dominated
by smooth disorder, so that the transport relaxation time is longer than the quantum
scattering time.

\begin{figure}[t]
    \includegraphics[width=1.00\columnwidth]{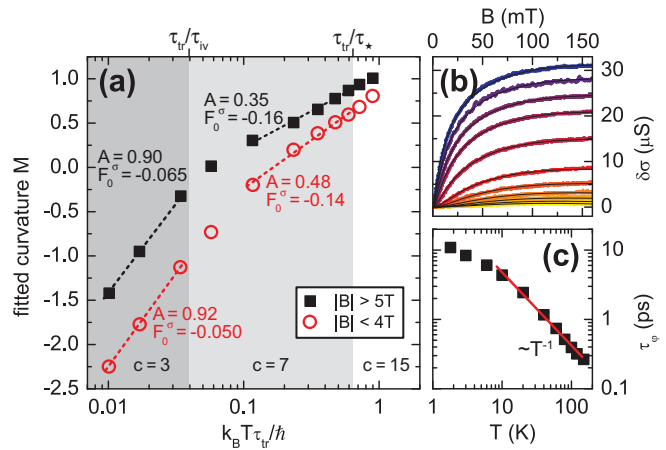}
    \caption{(a) The curvature $M$ of the parabolic MR extracted from the fits to the
          curves shown in Fig.\ \ref{fig:magnetoresistivity}(b) at weak
          (circles) and strong (squares) magnetic field. A kink is
          observed at $T = \hbar/\tau_{\text{iv}}k_{\text{B}} \approx 7$\,K as the
          number of contributing multiplet channels changes from $c = 3$ to $c = 7$.
          From the slopes $A$ the Fermi-liquid constant $F_0^\sigma$ is calculated for the different regions.
          (b) Weak localization correction to the conductivity at weak fields. Straight lines are fits of
          the graphene-specific model \cite{mccann-WL} that allows to determine the scattering
          times $\tau_\star$, $\tau_{\text{iv}}$ and the phase-coherence time $\tau_\varphi$.
          (c) As expected for electron-electron scattering in the diffusive regime $\tau_\varphi$
          decays as $T^{-1}$. \label{fig:fit-results}}
\end{figure}

At weak fields the WL
correction is clearly visible as a zero-field peak. We have characterized it carefully and
can accurately describe the anomaly within the standard graphene model
\cite{savchenko-WL, mccann-WL}.
The evaluation shown in Fig.\ \ref{fig:fit-results}(b) yields the
intravalley scattering time responsible for
the breaking of the effective time-reversal symmetry
 within a single valley $\tau_\star = 0.075$\,ps and the intervalley
 scattering time $\tau_{\text{iv}} = 1.1$\,ps, together with the phase
relaxation time $\tau_\varphi$, which is strongly temperature
 dependent [cf.\ Fig.\ \ref{fig:fit-results}(c)] and decays with
increasing $T$ roughly as $T^{-1}$ as expected for electron-electron scattering
in the diffusive regime \cite{savchenko-WL, altshuler}.
The WL data are qualitatively consistent
with findings in exfoliated graphene where the typical time scales are $\tau_{\text{iv}} = 0.8 \ldots 2.5$\,ps and $\tau_\star = 0.05 \ldots 0.3$\,ps \cite{savchenko-WL, horsell-UCF}. The deviation from the $T^{-1}$ dependence
at low $T$ presumably stems from the presence of magnetic impurities \cite{lara-avila-EEI}.

At stronger fields we observe that the curves have
roughly parabolic shape and vary with temperature,
showing an evolution from
negative to positive magnetoresistivity as the temperature rises.
Thus, the observed $\ln T$
dependence (see the analysis below) indicates that we are dealing with
the quantum contribution due to EEI as classical contributions to the magnetoresistance are temperature independent
at low $T$. Another striking feature of
Fig.~\ref{fig:magnetoresistivity}(a) is a crossing point of
low-$T$ curves that moves to a higher value of magnetic field at
higher temperatures.

For further analysis the data are symmetrized with respect to $B$ and
relative differences $(\rho-\rho_0)/\rho_0^2 \approx -\delta\sigma$ are considered. We choose $\rho_0 =
\rho(0.5\,\text{T})$ to eliminate the effect of the WL
correction at $|B|<0.5$\,T.
%
The resulting curves $(\rho-\rho_0)/\rho_0^2(B)$ are plotted in Fig.\
\ref{fig:magnetoresistivity}(b). Again the
transition from negative to positive magnetoresistance with
rising temperature is visible. A closer look at intermediate temperatures reveals
that the curvature differs in its sign even within the same curve between weaker ($|B|<4.5$\,T)
and stronger fields. Indeed for all curves two separate
parabolas $(\omega_{\text{c}}\tau_{\text{tr}})^2\frac{e^2}{2 \pi^2 \hbar}\cdot M$
with different curvatures $M_{\text{i}}$ and $M_{\text{o}}$ have to be taken into account for the
inner and the outer field region respectively, where $\omega_{\text{c}} = \frac{v_{\text{F}} e}{\hbar \sqrt{\pi n}}
B$ is the cyclotron frequency. As a particular
example Fig.\ \ref{fig:magnetoresistivity}(c) shows this decomposition
for the lowest ($T=1.8$\,K) and an intermediate ($T=20$\,K) temperature. The
transition field of $\approx 4.5$\,T is constant for all
temperatures.

We analyze the data with the help of a model that successfully
described the magnetoresistance in many 2DEGs in non-quantizing fields
(see Refs \cite{GMPRB} and \cite{woelfle-nonmarkovian} for
review) and is equally suited for graphene at large dimensionless conductance.
The model takes EEI into account and covers the crossover
from the diffusive to the ballistic regime. The diffusive
regime where $k_{\text{B}}T\tau_{\text{tr}}/\hbar \ll 1$ is observed for the lowest-temperature curves.
The transport time $\tau_{\text{tr}} \approx 0.045$\,ps is derived from the Drude resistivity.
The higher-$T$ curves correspond to the crossover regime, where the importance
of ballistic effects \cite{ZNA1,ZNA2,GMPRL,GMPRB} 
increases. Still, for all studied
temperatures and samples the condition $k_{\text{B}}T\tau_{\text{tr}}/\hbar \alt
1$ is fulfilled which means that the ballistic regime is not reached.

In the diffusive regime, EEI induces a logarithmic-in-$T$, $B$-independent
correction $\delta\sigma_{\text{ee}}$ to the longitudinal conductivity,
while not affecting the Hall conductivity \cite{altshuler}.
As a result, EEI gives rise to a magnetoresistance
$\delta\rho/\rho_0\propto (\mu B)^2\delta\sigma_{\text{ee}}/\sigma_0$ (here $\sigma_0$
is the zero-$B$ Drude conductivity)
which is parabolic in $B$ \cite{Houghton82,Girvin82}:
\begin{equation}
    \rho=\rho_0 + [(\omega_{\text{c}} \tau_{\text{tr}})^2 - 1]
        \frac{e^2 \rho^2_0}{2\pi^2 \hbar} \left[A \cdot
          \ln\left(\frac{k_{\text{B}} T
              \tau_{\text{tr}}}{\hbar}\right) \right]
        \text{,} \label{eq:EEI}
\end{equation}
This correction includes two contributions: the Coulomb term and the
Hartree term. The coefficient $A$ is the sum of these two terms
\begin{equation}
    A = A_{\text{C}} + A_{\text{H}} = 1 + c\left[1-\frac{\ln(1 +
            F_0^\sigma)}{F_0^\sigma}\right]\text{,} \label{eq:AF}
\end{equation}
with the Fermi liquid constant $F_0^\sigma$.
It is a measure for the interaction strength and in the case of weak interaction it
can be calculated within Thomas-Fermi approximation by integrating
the interaction matrix element
over all scattering angles \cite{shytov-EEI}:
\begin{equation}
F_{\text{0, theo}}^\sigma = - \alpha
\int{\frac{\cos^2\theta/2}{2\pi\left(\sin\theta/2 + 2\alpha\right)}\text{d}\theta}
\text{,} \label{eq:F0sigma}
\end{equation}
with the interaction constant $\alpha =
e_\ast^2/(4\pi\varepsilon_0\hbar v_{\text{F}}) \approx 0.41$ and an
effective charge $e_\ast = e \sqrt{2/(\varepsilon + 1)} \approx
0.43\,e$ taking into account the screening of Coulomb interaction by
the SiC substrate ($\varepsilon_{\text{SiC}} = 9.66$). For the given
case equations (\ref{eq:AF}) and (\ref{eq:F0sigma}) yield $A(F_{\text{0, theo}}^\sigma)
\approx 1-0.046 \cdot c$ and $F_{\text{0,
    theo}}^\sigma \approx -0.09$ respectively. The constant $c$ is  the number of multiplet
channels participating to EEI.

Here the peculiarity of graphene enters at the following two points.
First, the
$\cos^2\theta/2$ reflects the influence of chirality on
scattering of charge carriers in graphene.
Second, the channel statistics of graphene is different from
a standard 2DEG and depends on the interplay of symmetry-breaking time
scales. In our samples the hierarchy is $\tau_{\text{tr}} < \tau_\star
< \tau_{\text{iv}}$ as extracted from the analysis of the WL peak.
This hierarchy is also found
in exfoliated graphene \cite{savchenko-WL}. In perfect graphene the spin
and valley degeneracy gives rise to
$15$ multiplet channels.
For temperatures $T < \hbar/k_{\text{B}}\tau_\star$ the breaking of the time-reversal symmetry
within a single valley (due to, for example, the trigonal warping or
ripples) suppresses 8 channels out of 15, yielding $c = 7$
\cite{shytov-EEI}. The
intervalley scattering further reduces the number of multiplet
channels to $c=3$ (spin-triplet). This standard 2DEG case is expected for
sufficiently low temperatures $T < T_{\text{iv}} :=
\hbar/k_{\text{B}}\tau_{\text{iv}} \approx 7$\,K.

When analyzing the prefactors $A$ of formula (\ref{eq:EEI}), one can expect to see the predicted
crossover between $c=3$ and $c=7$ in the temperature dependence of the curvature of the magnetoresistivity.
Figure \ref{fig:fit-results} plots the curvatures $M_{\text{i}}$ for the inner (circles) and $M_{\text{o}}$
for the outer (squares) field region as a function of temperature (more explicitly $k_{\text{B}} T \tau_{\text{tr}}/\hbar$).
In this plot two regions of different slope can be distinguished, with a crossover at $T_{\text{iv}} \approx 7$\,K,
that perfectly fits the temperature scale given by $\tau_{\text{iv}}$ (extracted from WL data).
From Eq.\ (\ref{eq:EEI}) it is obvious that the slope of the curvatures $M$ in this plot over $k_{\text{B}} T \tau_{\text{tr}}/\hbar$
(on logarithmic scale) is equal to the coefficient $A$.
Hence the change in slope is caused by a different number of contributing multiplet channels in the two regimes.
Using $c=3$ as expected for $T < T_{\text{iv}}$ a Fermi liquid constant
of $F_{\text{0,i}}^\sigma = -0.05$ and $F_{\text{0,o}}^\sigma = -0.06$ can be extracted for
the inner and outer field region respectively. For $T > T_{\text{iv}}$ $F_{\text{0,i}}^\sigma = -0.14$
and $F_{\text{0,o}}^\sigma = -0.16$ are found with $c=7$. The obtained values are reasonably close
to the value of the theoretically expected $F_{\text{0,theo}}^\sigma = -0.09$.
The higher values at higher temperatures can qualitatively be attributed to the crossover towards
ballistics (see discussion below).

It is worthwhile to compare the extracted values of $F_{\text{0}}^\sigma$
to those measured in a 2D electron gas in conventional semiconductor structures.
The main factor determining the Fermi liquid constant is the
widely used parameter $r_\text{s}$ controlling the ratio of characteristic
potential and kinetic energies. It depends on the electron charge, the
dielectric constant, and the Fermi velocity (in our case $r_\text{s}=\sqrt{2}\alpha$). 
The central difference to semiconductors is that there the Fermi velocity depends strongly
on the electron concentration $n$, so that $r_\text{s}$ varies substantially
(typically in the range from 2 to 20, leading to a change of $F_{\text{0}}^\sigma$ between
0.15 and 0.7; see, e.g., Ref.\ \cite{galaktionov}).
In contrast, in graphene the velocity is roughly $n$-independent, and
$r_\text{s}$ remains relatively small. In addition, for given $r_\text{s}$ the
Fermi-liquid constant is somewhat suppressed in graphene due to
chiral character of carriers, see Eq.~(\ref{eq:F0sigma}).

\begin{figure}[t]
    \includegraphics[width=1.00\columnwidth]{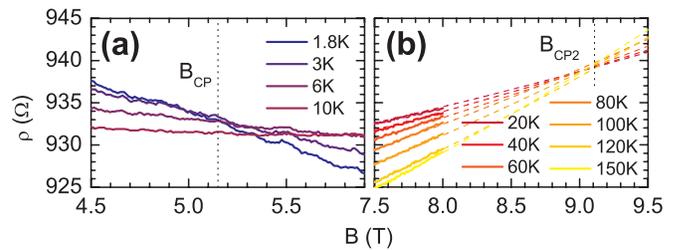}
    \caption{Zoom on the crossing points. (a) The crossing
          point at $B_{\text{CP}}=5.2$\,T is already visible in Fig.\
          \ref{fig:magnetoresistivity}. (b) For higher temperatures, an apparent
          crossing point at $B_{\text{CP2}} \simeq 9.1$\,T emerges by
          extrapolating (dashed lines) the parabolic shape of the
          curves onto $B>8$\,T. \label{fig:crossing-points}}
\end{figure}

A hallmark of the diffusive regime is that the curvature
of the MR correlates with the sign and amplitude of the correction to the zero-$B$
resistivity.
In particular this leads to the well defined crossing point (CP) in Fig.\ \ref{fig:magnetoresistivity}(a).
At this point at $B_{\text{CP}} = 5.2$\,T [Fig.\ \ref{fig:crossing-points}(a)] where $T$-dependent and $B$-dependent corrections exactly cancel,
the resistivity has the classical Drude value \cite{Minkov,Renard}. The value of this CP agrees well with $B_{\text{CP}} = 4.8$\,T
expected from $\omega_{\text{c}}\tau_{\text{tr}}=1$.
The overall consistency of our findings (together with the emergence of the crossing point)
confirms that the applied diffusive model
describes the physics properly and hence is a reliable way to separate the effect of EEI from other
corrections.

Now we turn to the observations which are beyond the
 purely diffusive model described by Eq.\ (\ref{eq:EEI}).
(I) First, we find different curvatures of the MR for
weak and strong magnetic fields [cf.\ Fig.\ \ref{fig:magnetoresistivity}(c)].
(II) Second, the higher-$T$ MR curves have a tendency to form an apparent
second crossing point [cf.\ Fig.\ \ref{fig:crossing-points}(b)].
These two observations suggest that the diffusive mechanism, which governs
the logarithmic temperature dependence of the curvature of the MR in the whole
range of magnetic fields and temperatures explored in this work, 
is supplemented by ballistic effects that become more pronounced at higher $T$ and
magnetic field.

Observation (I)
implies that the MR contains a $T$-independent contribution
which stems from ballistic scales \cite{GMPRL,GMPRB} and is sensitive
to the relation between $\omega_\text{c}$ and $\tau_{\text{tr}}$. 
As the magnetic field increases, for $B>1/\mu$ (which corresponds to
$\omega_{\text{c}}\tau_{\text{tr}}>1$) electrons can accomplish closed
cyclotron orbits which leads to an enhanced probability of scattering
angles $\theta \approx 0$ [sketch in Fig.\
\ref{fig:magnetoresistivity}(c)]. This correlation causes a change of
the strength of EEI [see Eq.\ (\ref{eq:F0sigma})]
in the ballistic contribution
and hence the overall curvature of the MR.
In addition, there might be a temperature-independent
contribution due to a classical mechanism of
magnetoresistance (due to non-Markovian processes, for review see
Ref.\ \cite{woelfle-nonmarkovian}) operative
in classically strong magnetic fields $B>1/\mu$ that leads to a vertical shift
between the values of $M_{\text{i}}$ and $M_{\text{o}}$.

Observation (II)
indicates that the $T$ dependence of the zero-$B$ resistivity $\rho_0$ does not fully determine
the curvature of the MR \cite{GMPRL,GMPRB,Renard,savchenko-MR}.
This signals the onset of a crossover to the ballistic regime at $T \geq 20$\,K.
Indeed, at $k_{\text{B}} T \tau_{\text{tr}} / \hbar < 1$ the $T$ dependence of $\rho_0$ stems from 
the combination of a logarithmic diffusive correction and a $T$-dependent correction $\delta\mu(T)$
to the mobility (Ballistic return processes may give rise to ballistic $T$-dependent corrections to the mobility \cite{cheianov}). At lowest temperatures, the logarithmic diffusive correction dominates.
With increasing $T$, the diffusive term decreases while $\delta\mu(T)$ grows.
Importantly, the processes giving rise to a $T$-dependent correction to the mobility
do not lead to the MR at finite $B$, in contrast to the logarithmic diffusive term. 
Therefore, $\delta\mu(T)$ only shifts all resistivity curves $\rho(B)$ in the vertical direction
(in the crossover to the ballistic regime the diffusive and ballistic contributions to this shift
become comparable), while the $T$ dependence of the MR curvature is still given solely by the logarithmic term. 
This explains the shift of the CP
towards an apparent high-$T$ CP at $B_{\text{CP2}} \simeq 9.1$\,T that emerges when
the MR parabolas are extrapolated to stronger fields [Fig.\ \ref{fig:crossing-points}(b)].
Its observation thus proves that ballistic corrections to $\rho_0$ gain importance
at higher temperatures.

Theoretical results for MR in the ballistic (and crossover)
regimes depend on the microscopic model of disorder
\cite{GMPRB,woelfle-nonmarkovian}.
Therefore, a systematic study of MR in
the crossover and ballistic regimes may be helpful for clarifying the
character of disorder in the system (which is particularly important
for electron transport in graphene \cite{OGM06}); we relegate this to a future
work. In this paper, we restrict ourselves to applying the diffusive
model supplemented at strong $B$ by a $T$-independent ballistic
contribution, which is a reasonably good approximation in the
considered temperature range.

In conclusion, we have studied the magnetotransport in graphene epitaxially grown on silicon carbide.
We observe a temperature-dependent parabolic magnetoresistivity which is a manifestation of the EEI.
The temperature dependence of the MR curvature is logarithmic, as expected for the diffusive regime.
We find a pronounced kink in this dependence, that is related to the reduction of the number of multiplet
modes contributing to EEI from 7 to 3 due to intervalley scattering.

We show that our model of diffusive transport with ballistic contributions can describe
the underlying physics properly.
The  analysis of the temperature dependence of the magnetoresistance
allows for an accurate description of the EEI interaction correction in the
diffusive regime.
At magnetic fields
higher than the observed crossing point ($\omega_\text{c} \tau_{tr} > 1$) the magnetoresistance shows
an additional (compared to weak field) $T$-independent parabolic contribution due to ballistic effects.
Further quantitative
analysis of the crossover and ballistic regimes should be helpful for shedding light on the
nature of disorder in graphene structures.

Our work demonstrates that the analysis of MR in
the intermediate range of $B$ is a convenient tool for exploring many-body effects in graphene.
In particular, this method allows to precisely distinguish the effect of EEI from additional
corrections to the temperature dependent resistivity that also have logarithmic temperature
dependence. Furthermore, our approach allowed us to determine the Coulomb interaction strength
in our samples with a rather high accuracy.
The value of the interaction constant obtained from the fit to the experimental data is
$F_0^\sigma \approx -0.06$ at low $T$ and $-0.15$ at high $T$ (where the system approaches
the ballistic regime), which is close to the theoretical estimate for graphene on silicon
carbide, $F_0^\sigma \approx -0.09$.
These findings should also directly apply
to other graphene samples (e.g., to exfoliated graphene) provided the system sizes are sufficiently large.

\begin{acknowledgments}
We gratefully acknowledge support within the Cluster of Excellence {\it
Engineering of Advanced Materials} (www.eam.uni-erlangen.de), DFG
Center for Functional Nanostructures, and DFG SPP 1459 ``Graphene''. We acknowledge the use of the hydrogen intercalation furnace provided by Th.\ Seyller.
\end{acknowledgments}

\bibliography{bibliography-EEI}

\clearpage
\appendix

\section{Supplemental Material for \lq Manifestation of electron-electron interaction in the
  magnetoresistance of graphene\rq}
This supplemental material is dedicated to elucidate the generality and reproducibility of the findings reported in the main text. We have observed the parabolic shape of the magnetoresistance (MR) in between the weak-localization (WL) regime and the Landau quantization regime and its logarithmic temperature dependence for many samples. We have investigated it thoroughly for four samples. We decided to show data of sample S1 as one consistent data set in the main text. The data of the remaining three samples (S2, S3 and S4) are subject to this supplemental material.  
The quasi free-standing monolayer graphene (QFMLG) sample S2 had smaller size than S1 which made the analysis of the low temperature
behavior difficult and increased the amplitude of the temperature-independent quasiclassical MR.
The monolayer samples S3 and S4 (both not QFMLG) showed a large phonon contribution to the MR
in view of a strong coupling to the substrate. Therefore, sample S1 was best suited for
our analysis of electron-electron interaction (EEI) effects. Nevertheless, samples S2, S3, and S4 show 
the EEI effects in the temperature dependence of the MR 
quantitatively consistent with those in sample S1 (details are presented below).
\subsection{Similar sample type}
Figure \ref{fig:SM-H21-10} displays results of QFMLG sample S2 that is similar to sample S1 presented in the main text. The main difference is that S2 is smaller (10\,$\mu$m wide, 70\,$\mu$m long) which made it difficult to analyze the low temperature behavior and increased the amplitude of the $T$-independent quasiclassical MR. In particular at lowest temperatures universal conductance fluctuations become apparent. Along with the somewhat lower mobility, the crossing point has moved to higher fields, and a field driven cross-over can not be detected. The evaluation of the curvature of the MR, however, shows the same cross-over from $c=3$ to $c=7$ with rising temperature. The analysis yields similar values of the Fermi-liquid constant $F_0^\sigma$, at least for high temperatures. At low temperatures the data quality and density is unfortunately not as good as for sample S1 and therefore does not allow to extract $F_0^\sigma$ for the $c = 3$ region.    
\subsection{Different epitaxial graphene material}
Beyond QFMLG samples S1 and S2 presented in the main text and above, we investigated two samples S3 and 
\begin{figure}[H]
	\centering
    \includegraphics[width=0.95\columnwidth]{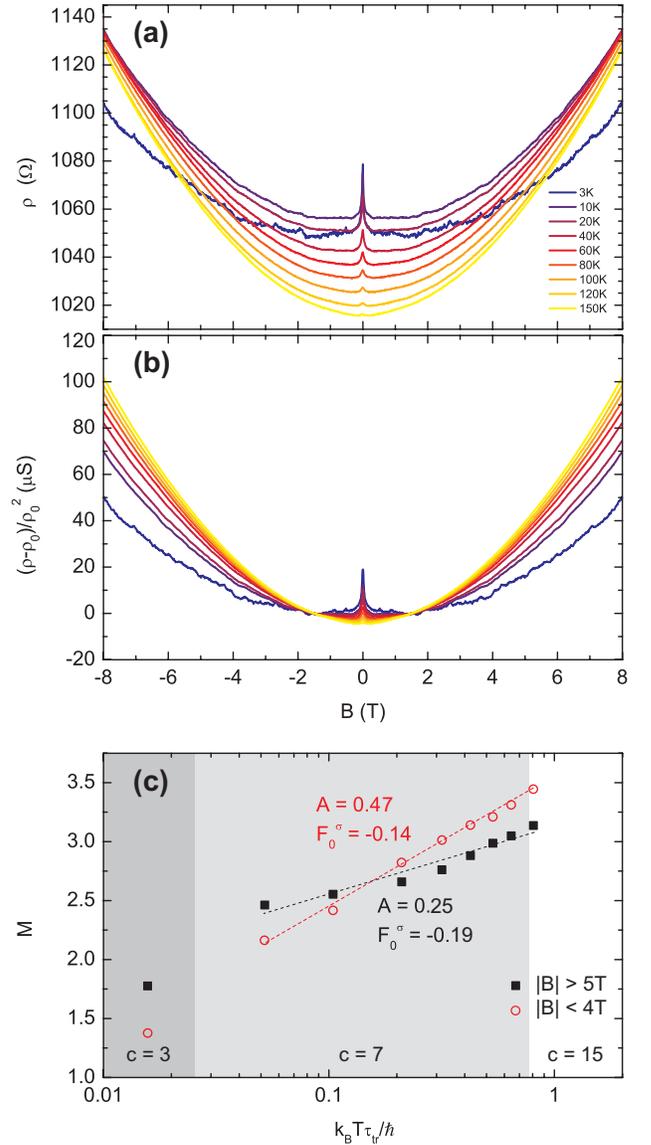}
    \caption{(a) The magnetoresistivity raw data exhibits a parabolic shape and an apparent crossing point at $\approx 9$\,T. (b) The relative magnetoresistivity shows the evolution of the curvatures with temperature. (c) Evaluation of the curvatures $M$ yields the Fermi-liquid constant. Intervalley scattering time $\tau_{iv} = 1.5$\,ps and intravalley scattering time $\tau_{\star} = 0.051$\,ps were extracted from WL data. \label{fig:SM-H21-10}}
\end{figure}
\begin{figure*}[ht]
    \includegraphics[width=0.98\textwidth]{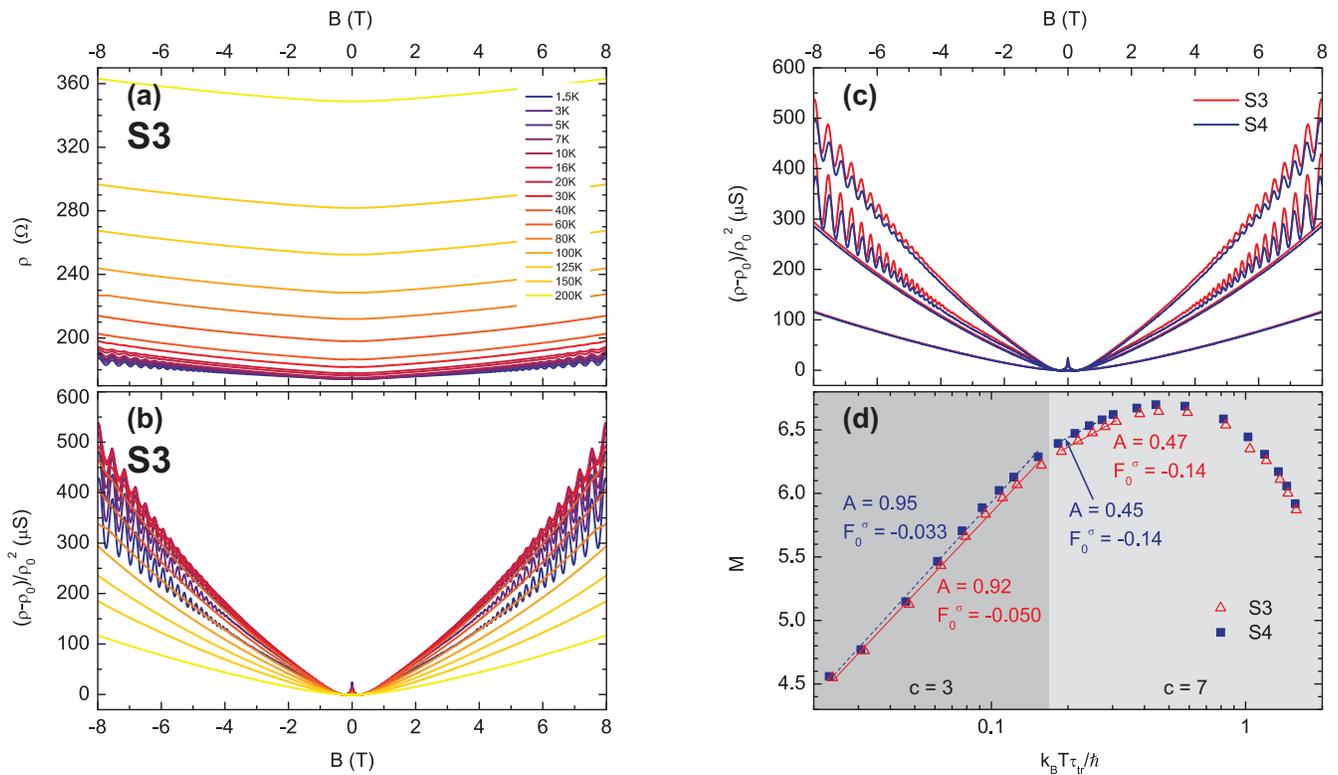}
    \caption{The magnetoresistivity raw data of sample S3 (a) and S4 shows a pronounced vertical shift due to the coupling to substrate phonons in addition to the parabolic MR. At low temperatures and high fields Shubnikov-de Haas oscillations are visible. The relative magnetoresistivity (b) displays the evolution of the MR curvature with temperature. As MLG samples S3 and S4 are processed on the same chip their electronic properties are virtually identical as illustrated in (c) where MR curves for 10\,K, 1.5\,K, 100\,K and 200\,K (from top to bottom) are compared. Thus, also the evaluation of the MR curvatures $M$ of samples S3 and S4 are nearly equal (d). Note that the extracted values are very similar to the findings on sample S1. The scattering times $\tau_{iv} = 0.53$\,ps and $\tau_{\star} = 0.025$\,ps were extracted from WL data. \label{fig:SM-SPP_MLG3_HB1and2}}
\end{figure*}  
S4 using monolayer graphene (MLG) \cite{emtsev-natmat}. Microscopically the materials differ in an interface layer underneath
the graphene \cite{riedl-QFMLG, speck-QFMLG}. Phenomenologically, they differ in the sign of charge carriers (holes for QFMLG and electrons
 for MLG), and in different coupling to substrate phonons \cite{hibino-SPP}. As a consequence of the latter, MLG has a substantial dependence of the resistivity $\rho$ on temperature $T$. It can readily be seen [Fig.\ \ref{fig:SM-SPP_MLG3_HB1and2}(a)] that a strong temperature dependent offset of the MR curves result. Therefore the occurrence of a crossing point is obscured. Figures \ref{fig:SM-SPP_MLG3_HB1and2}(b) and (d) show the further analysis carried out in exactly the same manner as presented in the main text. The enhanced mobility has two main consequences: (i) quantum oscillations occur at substantially lower fields as in S1. (ii) the obscured cross-over point also moves to lower fields and therefore the field range of the inner parabola (between WL regime and the obscured crossing-point) is compressed. Hence it is difficult to evaluate the inner curvature $M_i$ and we restrict ourselves to $M_o$. The evaluation of the Fermi-liquid constant results in excellent agreement with the values extracted for S1. This displays the robustness of the findings taking into account that a different graphene material, a different charge carrier density and a different charge carrier type is present. The downturn in Fig.\ \ref{fig:SM-SPP_MLG3_HB1and2}(d) happens close to $k_B T \tau_{tr}/\hbar = 1$ where the diffusive picture becomes inadequate. In addition, the phonon effects at high temperatures may
affect the $T$ dependence of the MR curvature.   
%


%
\subsection{Reproducibility}
One advantage of epitaxial graphene is the ability to very well reproduce data. As an example we show the MR of two samples S3 and S4 with nominally identical geometry (50\,$\mu$m wide, 330\,$\mu$m long) being processed on the same chip. Differences can hardly be recognized [Fig.\ \ref{fig:SM-SPP_MLG3_HB1and2}(c) and (d)]. The above examples, however, show that when different fabrication parameters or different processing are chosen, the resulting samples may differ, most evidently in mobility and/or charge density.

\end{document}